\documentclass [12pt,epsf] {article}
\usepackage{axodraw}

\catcode`@=12
\newcommand{\beq}{\begin{equation}}
\newcommand{\eeq}{\end{equation}}

\newcommand{\bea}{\begin{eqnarray}}
\newcommand{\eea}{\end{eqnarray}}

\begin{document}
\thispagestyle{empty}
\begin{flushright} UCRHEP-T341\\
OSU-HEP-02-07\\IFIC/02-26\\June 2002\
\end{flushright}
\vspace{0.5in}
\begin{center}
{\Large \bf Underlying $A_4$ Symmetry for the Neutrino\\ 
Mass Matrix and the Quark Mixing Matrix\\}
\vspace{1.5in}
{\bf K. S. Babu$^1$, Ernest Ma$^2$, and J. W. F. Valle$^3$\\}
\vspace{0.2in}
{\sl $^1$ Physics Department, Oklahoma State University, Stillwater,
Oklahoma 74078, USA\\}
{\sl $^2$ Physics Department, University of California, Riverside,
California 92521, USA\\}
{\sl $^3$ Instituto de F\'{\i}sica Corpuscular -- C.S.I.C.,
Universitat de Val{\`e}ncia,\\
Edificio Institutos, Aptdo.\ 22085, E--46071 Val{\`e}ncia, Spain\\}
\vspace{1in}
\end{center}
\begin{abstract}
The discrete non-Abelian symmetry $A_4$, valid at some high-energy scale, 
naturally leads to degenerate neutrino masses, without spoiling the hierarchy 
of charged-lepton masses.  Realistic neutrino mass splittings and mixing 
angles (one of which is necessarily maximal and the other large) are then 
induced radiatively in the context of softly broken supersymmetry.  The 
quark mixing matrix is also calculable in a similar way.  The mixing parameter 
$U_{e3}$ is predicted to be imaginary, leading to maximal CP violation in 
neutrino oscillations.  Neutrinoless double beta decay and $\tau \to \mu 
\gamma$ should be in the experimentally accessible range.
\end{abstract}

\newpage
\baselineskip 24pt

It has often be said that the mixing pattern of neutrinos, which
involves large angles, as evidenced by the atmospheric \cite{atm} and
solar \cite{sol} neutrino data, is unexpected and difficult to
understand, given that the quark charged-current mixing matrix
$V_{CKM}$ involves only small angles.  However, as shown below, both
can be explained in a simple and unified way as small radiative
corrections of a fixed pattern, valid at some high-energy scale as the
result of an underlying symmetry, which we identify here as $A_4$, the
non-Abelian discrete symmetry group of the tetrahedron \cite{a4}.  We
show that at the high scale, neutrino masses are degenerate and
$V_{CKM}$ is the identity matrix.  We then calculate the radiative
corrections down at the electroweak scale in the framework of softly
broken supersymmetry \cite{Chankowski:2000fp,bdm} and obtain realistic
versions of ${\cal M}_\nu$ and $V_{CKM}$.  The reason that neutrino
mixing involves large angles is a simple consequence of degenerate
perturbation theory, where a small off-diagonal term induces maximal
mixing between two states of equal energy, whereas in the quark
sector with hierarchical masses, the same small off-diagonal element
induces only a small mixing.

Our starting point is the model of Ref.~\cite{a4}, but with the
following two important improvements.  (I) Instead of breaking
$A_4$ spontaneously at the electroweak scale, it is now broken at
a very high scale.  (II) Supersymmetry is added with explicit soft
breaking terms which also break $A_4$.  The resulting theory at
the electroweak scale is a specific version of the MSSM (Minimal
Supersymmetric Standard Model), where the scalar lepton and quark
sectors are correlated with ${\cal M}_\nu$ and $V_{CKM}$. In this way
we also provide a theoretical framework for realizing the neutrino
unification idea suggested in the first paper of Ref.
\cite{Chankowski:2000fp}, but with different specific predictions.

The non-Abelian discrete finite group $A_4$ consists of 12 elements 
and has 4 irreducible
representations.  Three are one-dimensional, \underline {1},
\underline {1}$'$, \underline {1}$''$, and one is three-dimensional,
\underline {3}, with the decomposition
\begin{equation}
\underline {3} \times \underline {3} = \underline {1} + \underline {1}' +
\underline {1}'' + \underline {3} + \underline {3}.
\end{equation}
The usual quark, lepton, and Higgs superfields transform under $A_4$
as follows:
\begin{eqnarray}
&& \hat Q_i = (\hat u_i, \hat d_i), ~\hat L_i = (\hat \nu_i, \hat e_i) \sim
\underline {3}, ~~~\hat \phi_{1,2} \sim \underline {1}, \\
&& \hat u^c_1, ~\hat d^c_1, ~\hat e^c_1 \sim \underline {1}, ~~~ \hat u^c_2,
~\hat d^c_2, ~\hat e^c_2 \sim \underline {1}', ~~~ \hat u^c_3, ~\hat d^c_3,
~\hat e^c_3 \sim \underline {1}''.
\end{eqnarray}
We then add the following heavy quark, lepton, and Higgs superfields:
\begin{equation}
\hat U_i, ~\hat U^c_i, ~\hat D_i, ~\hat D^c_i, ~\hat E_i, ~\hat E^c_i,
~\hat N^c_i, ~\hat \chi_i \sim \underline {3},
\end{equation}
which are all $SU(2)$ singlets.  The superpotential of this model is then
given by
\begin{eqnarray}
\hat W &=& M_U \hat U_i \hat U^c_i + f_u \hat Q_i \hat U^c_i \hat \phi_2 +
h^u_{ijk} \hat U_i \hat u^c_j \hat \chi_k \nonumber \\
&+& M_D \hat D_i \hat D^c_i + f_d \hat Q_i \hat D^c_i \hat \phi_1 +
h^d_{ijk} \hat D_i \hat d^c_j \hat \chi_k \nonumber \\
&+& M_E \hat E_i \hat E^c_i + f_e \hat L_i \hat E^c_i \hat \phi_1 +
h^e_{ijk} \hat E_i \hat e^c_j \hat \chi_k \nonumber \\
&+& {1 \over 2} M_N \hat N^c_i \hat N^c_i + f_N \hat L_i \hat N^c_i
\hat \phi_2 + \mu \hat \phi_1 \hat \phi_2 \nonumber \\
&+& {1 \over 2} M_\chi \hat \chi_i \hat \chi_i + h_\chi \hat \chi_1
\hat \chi_2 \hat \chi_3,
\end{eqnarray}
where we have adopted the usual assignment of $R$ parity to
distinguish between the Higgs superfields, i.e. $\hat \phi_{1,2}$ and
$\hat \chi_i$, from the quark and lepton superfields.  We have also
forbidden the terms $\hat \chi_i \hat N^c_j \hat N^c_k$, etc. by
assigning
\begin{equation}
\hat \chi_i \sim \omega, ~~~\hat u^c_i, ~\hat d^c_i, ~\hat e^c_i \sim \omega^2,
\end{equation}
and all others $\sim$ 1 under a separate discrete symmetry $Z_3$ (with
$\omega^3 = 1$ and $1 + \omega + \omega^2 = 0$).  However, $Z_3$ is
allowed to be broken explicitly but only softly, which is uniquely
accomplished in the above by $M_\chi \neq 0$.

The scalar potential involving $\chi_i$ is given by
\begin{equation}
V = |M_\chi \chi_1 + h_\chi \chi_2 \chi_3|^2 + |M_\chi \chi_2 + h_\chi \chi_3
\chi_1|^2 + |M_\chi \chi_3 + h_\chi \chi_1 \chi_2|^2,
\end{equation}
which has the supersymmetric solution $(V=0)$
\begin{equation}
\langle \chi_1 \rangle = \langle \chi_2 \rangle = \langle \chi_3 \rangle =
u = -M_\chi/h_\chi,
\end{equation}
so that the breaking of $A_4$ at the high scale $M_\chi$ does not
break the supersymmetry.  [Note that Eq.~(8) is only possible because
$A_4$ allows the invariant symmetric product of \underline {3}
$\times$ \underline {3} $\times$ \underline {3}, a highly nontrivial 
property not shared for example 
by the triplet representation of either SO(3) or SU(3).]

Consider now the $6 \times 6$ Dirac mass matrix linking $(e_i,E_i)$ to
$(e_j^c,E_j^c)$.
\begin{equation}
{\cal M}_{eE} = \left[ \begin{array} {c@{\quad}c@{\quad}c@{\quad}c@{\quad}c@
{\quad}c} 0 & 0 & 0 & f_e v_1 & 0 & 0 \\ 0 & 0 & 0 & 0 & f_e v_1 & 0 \\
0 & 0 & 0 & 0 & 0 & f_e v_1 \\ h_1^e u & h_2^e u & h_3^e u & M_E & 0 & 0 \\
h_1^e u & h_2^e \omega u & h_3^e \omega^2 u & 0 & M_E & 0 \\
h_1^e u & h_2^e \omega^2 u & h_3^e \omega u & 0 & 0 & M_E \end{array} \right],
\end{equation}
where $v_1 = \langle \phi_1^0 \rangle$, and Eq.~(17) of the first
paper of Ref.~\cite{a4} has been used, with similar forms for the
quark mass matrices. The reduced $3 \times 3$ Dirac mass matrix for
the charged leptons is then
\begin{equation}
{\cal M}_e = U_L \left[ \begin{array} {c@{\quad}c@{\quad}c} {h_1^e}' & 0 & 0
\\ 0 & {h_2^e}' & 0 \\ 0 & 0 & {h_3^e}' \end{array} \right] {\sqrt 3 f_e v_1
u \over M_E},
\end{equation}
where ${h_i^e}' = h_i^e [1+(h_i^e u)^2/M_E^2]^{-1/2}$ and
\begin{equation}
U_L = {1 \over \sqrt 3} \left[ \begin{array} {c@{\quad}c@{\quad}c} 1 & 1 & 1
\\ 1 & \omega & \omega^2 \\ 1 & \omega^2 & \omega \end{array} \right].
\end{equation}
This shows that charged-lepton masses are allowed to be all different,
despite the imposition of the $A_4$ symmetry, because there exist
three inequivalent one-dimensional representations.  [Note that the
permutation symmetry groups $S_3$ and $S_4$ have only two inequivalent
one-dimensional representations and $S_3$ has no three-dimensional
representation.]  Clearly, the $up$ and $down$ quark mass matrices are
obtained in the same way, with the important conclusion that the
charged-current mixing matrix $V_{CKM}$ is automatically equal to the
identity matrix, because both are diagonalized by $U_L$.  Corrections
to $V_{CKM} = 1$ may then be ascribed to the structure of the soft
supersymmetry breaking sector \cite{bdm,Gabbiani:1996hi}.

In the neutrino sector, the $6 \times 6$ Majorana mass matrix spanning
$(\nu_e, \nu_\mu, \nu_\tau, N_1^c, N_2^c, N_3^c)$ is given by
\begin{equation}
{\cal M}_{\nu N} = \left[ \begin{array} {c@{\quad}c} 0 & U_L f_N v_2
\\ U_L^T f_N v_2 & M_N \end{array} \right],
\end{equation}
where $v_2 = \langle \phi_2^0 \rangle$.  Hence the $3 \times 3$ seesaw mass
matrix for $(\nu_e, \nu_\mu, \nu_\tau)$ becomes
\begin{equation}
{\cal M}_\nu = {f_N^2 v_2^2 \over M_N} U_L^T U_L = {f_N^2 v_2^2 \over M_N}
\left[ \begin{array} {c@{\quad}c@{\quad}c} 1 & 0 & 0 \\ 0 & 0 & 1 \\
0 & 1 & 0 \end{array} \right].
\end{equation}
This shows that neutrino masses are degenerate at this stage.

Consider now the above as coming from an effective dimension-five operator
\cite{eff}
\begin{equation}
{f_N^2 \over M_N} \lambda_{ij} \nu_i \nu_j \phi_2^0 \phi_2^0,
\end{equation}
where $\lambda_{ee} = \lambda_{\mu \tau} = \lambda_{\tau \mu} = 1$ and
all other $\lambda$'s are zero, which is valid at some high scale.  As
we come down to the electroweak scale, Eq.~(14) is corrected
\cite{blp} by the wavefunction renormalizations of $\nu_e$, $\nu_\mu$,
and $\nu_\tau$, as well as the corresponding vertex renormalizations.
Even if only the standard model is considered, this will lift the
degeneracy of Eq.~(13) because of the different charged-lepton masses.
The resulting pattern, i.e.  $\lambda_{ee} = 1 + 2 m_e^2 \epsilon$,
$\lambda_{\mu \tau} = \lambda_{\tau \mu} = 1 + (m_\mu^2 + m_\tau^2)
\epsilon$, where $\epsilon \sim 1/(16 \pi^2 v^2)~{\rm ln}(M_N/M_Z)$,
is however not suitable for explaining the present data on neutrino
oscillations.  On the other hand, other radiative corrections exist in
the context of softly broken supersymmetry with a general slepton mass
matrix \cite{Chankowski:2000fp}.  Given the structure of
$\lambda_{ij}$ at the high scale, its form at the low scale is necessarily 
fixed to first order as
\begin{equation}
\lambda_{ij} = \left[ \begin{array} {c@{\quad}c@{\quad}c} 1 + 2 \delta_{ee} &
\delta_{e \mu} + \delta_{e \tau} & \delta_{e \mu} + \delta_{e \tau} \\
\delta_{e \mu} + \delta_{e \tau} & 2 \delta_{\mu \tau} & 1 + \delta_{\mu \mu}
+ \delta_{\tau \tau} \\ \delta_{e \mu} + \delta_{e \tau} & 1 + \delta_{\mu \mu}
+ \delta_{\tau \tau} & 2 \delta_{\mu \tau} \end{array} \right],
\end{equation}
where we have assumed all parameters to be real as a first approximation.  
[The above matrix is obtained by multiplying that of Eq.~(13) on the left 
and on the right by all possible $\nu_i \to \nu_j$ transitions.]  Let us 
rewrite the above with $\delta_0 \equiv
\delta_{\mu \mu} + \delta_{\tau \tau} - 2 \delta_{\mu \tau}$, $\delta
\equiv 2 \delta_{\mu \tau}$, $\delta' \equiv \delta_{ee} - \delta_{\mu
  \mu}/2 - \delta_{\tau \tau}/2 - \delta_{\mu \tau}$, and $\delta''
\equiv \delta_{e \mu} + \delta_{e \tau}$.  Then
\begin{equation}
\lambda_{ij} = \left[ \begin{array} {c@{\quad}c@{\quad}c} 1 + \delta_0 +
2 \delta + 2 \delta' & \delta'' & \delta'' \\ \delta'' & \delta & 1 +
\delta_0 + \delta \\ \delta'' & 1 + \delta_0 + \delta & \delta \end{array}
\right],
\end{equation}
and the $exact$ eigenvectors and eigenvalues are easily obtained:
\begin{equation}
\left[ \begin{array} {c} \nu_1 \\ \nu_2 \\ \nu_3 \end{array} \right] =
\left[ \begin{array} {c@{\quad}c@{\quad}c} \cos \theta & \sin \theta/\sqrt 2
& \sin \theta/\sqrt 2 \\ -\sin \theta & \cos \theta/\sqrt 2 & \cos \theta/
\sqrt 2 \\ 0 & -1/\sqrt 2 & 1/\sqrt 2 \end{array} \right] \left[
\begin{array} {c} \nu_e \\ \nu_\mu \\ \nu_\tau \end{array} \right],
\end{equation}
and
\begin{eqnarray}
\lambda_1 &=& 1 + \delta_0 + 2 \delta + \delta' - \sqrt{\delta'^2 + 2
\delta''^2}, \\
\lambda_2 &=& 1 + \delta_0 + 2 \delta + \delta' + \sqrt{\delta'^2 + 2
\delta''^2}, \\
\lambda_3 &=& -1 - \delta_0.
\end{eqnarray}
With $\delta''^2/\delta'^2$ of order unity, this is then a very satisfactory
description of present neutrino-oscillation data, i.e.
\begin{equation}
\sin^2 2 \theta_{atm} = 1, ~~~ \tan^2 \theta_{sol} = {\delta''^2 \over
\delta''^2 + \delta'^2 - \delta' \sqrt {\delta'^2 + 2\delta''^2}} = 0.44,
\end{equation}
if $\delta' < 0$ and $|\delta''/\delta'| = 1.7$.  Note that for
$\delta'' = \delta'$ Eq.~(16) reproduces that proposed in the second
paper of Ref.~\cite{a4}.  Whereas the latter was simply an ansatz, the
form of Eq.~(16) here is a $necessary$ consequence of our assumption
that radiative corrections are responsible for the splitting of the
neutrino mass degeneracy enforced by the discrete $A_4$ symmetry.
Assuming that $\delta', \delta'' << \delta$, we now have
\begin{equation}
\Delta m^2_{31} \simeq \Delta m^2_{32} \simeq 4 \delta m_0^2, ~~~
\Delta m^2_{12} \simeq 4 \sqrt{\delta'^2 + 2 \delta''^2} m_0^2,
\end{equation}
where $m_0$ is the common mass of all 3 neutrinos.

Note that $U_{e3} = 0$ in Eq.~(17), which would imply the absence of $CP$
violation in neutrino oscillations.  However, if we do not assume
$\lambda_{ij}$ to be real, then it has one complex phase which cannot be
rotated away.  Without loss of generality, we now rewrite Eq.~(16) as
\begin{equation}
\lambda_{ij} = \left[ \begin{array} {c@{\quad}c@{\quad}c} 1 +
2 \delta + 2 \delta' & \delta'' & \delta''^* \\ \delta'' & \delta & 1
+ \delta \\ \delta''^* & 1 + \delta & \delta \end{array}
\right],
\end{equation}
where we have redefined $1 + \delta_0$ as 1, and $\delta$, $\delta'$
are real.  Although this mass matrix cannot be diagonalized exactly,
if we assume that $\delta'$, $Re \delta''$ and $(Im
\delta'')^2/\delta$ are all much smaller than $\delta$ in magnitude,
then Eqs.~(17) to (22) are again valid (but only approximately) with
the following changes:
\begin{equation}
U_{e3} = {i Im \delta'' \over \sqrt 2 \delta}, ~~~ \delta' \to \delta' +
{(Im \delta'')^2 \over 2 \delta}, ~~~ \delta'' \to Re \delta''.
\end{equation}
Note the important result that $U_{e3}$ is imaginary.  Thus $CP$
violation is $predicted$ to be $maximal$ in this model for neutrino
oscillations.  Using Eq.~(22), we also have the relationship
\begin{equation}
\left[ {\Delta m_{12}^2 \over \Delta m_{32}^2} \right]^2 \simeq \left[
{\delta' \over \delta} + |U_{e3}|^2 \right]^2 + \left[ {Re \delta'' \over
\delta} \right]^2.
\end{equation}
Note that $|U_{e3}|$ is naturally of the order $|\Delta m^2_{12}/\Delta 
m^2_{32}|^{1/2} \simeq 0.14$.  This result depends only on the form of 
Eq.~(23), which is itself derived in the most general way.

It remains to be shown in the rest of this paper that realistic values
of $\delta$, $\delta'$, and $\delta''$ are possible from the soft
breaking of supersymmetry, without running into conflict with present
limits on neutrinoless double beta ($\beta\beta_{0\nu}$) decay 
and lepton flavor violating processes such as $\tau \to \mu \gamma$.

Let us calculate $\delta$ in the context of supersymmetry.  We show in
Figures 1 and 2 the wavefunction and vertex corrections respectively
due to $\tilde \mu_L - \tilde \tau_L$ mixing.  Let the two scalar mass
eigenstates have masses $\tilde m_{1,2}$ and their mixing angle be
$\theta$.  For illustration, let us take the approximation that
$\tilde m_1^2 >> \mu^2 >> M_{1,2}^2 = \tilde m_2^2$, where $\mu$ is
the superpotential Higgsino mixing term, while $M_{1,2}$ denote the
soft supersymmetry breaking gaugino mass parameters.  We then obtain
\begin{equation}
\delta \simeq {\sin \theta \cos \theta \over 16 \pi^2} \left[ (3 g_2^2 -
g_1^2) \ln {\tilde m_1^2 \over \mu^2} - {1 \over 4} (3 g_2^2 + g_1^2) \left(
\ln {\tilde m_1^2 \over \tilde m_2^2} - {1 \over 2} \right) \right] .
\end{equation}
Using Eq.~(22) and taking $\Delta m^2_{32} = 2.5 \times 10^{-3}$ eV$^2$ from
atmospheric neutrino oscillations, we find $\delta = 3.9 \times 10^{-3}
(0.4~{\rm eV}/m_0)^2$.  This implies that
\begin{equation}
\left[ \ln {\tilde m_1^2 \over \mu^2} - 0.3 \left( \ln {\tilde m_1^2 \over
\tilde m_2^2} - {1 \over 2} \right) \right] \sin \theta \cos \theta
\simeq 0.535 \left( {0.4~{\rm eV} \over m_0} \right)^2.
\end{equation}
To the extent that this factor cannot be much greater than one, the
common mass $m_0$ as probed in neutrinoless double beta decay
\cite{bb} cannot be much lower than the present upper bound of about
0.4 eV. This is in sharp contrast to the scenario proposed in the first paper 
of Ref.~\cite{Chankowski:2000fp} where $\beta\beta_{0\nu}$ decay is strongly
suppressed.

Similarly, $\delta'' = \delta_{e \mu} + \delta_{\tau e} = \delta_{e
  \mu} + \delta^*_{e \tau}$ is determined by $\tilde e_L - \tilde
\mu_L$ and $\tilde e_L - \tilde \tau_L$ mixing.  Using the
experimental bound of $|U_{e3}| < 0.16$, we find $Im \delta'' < 8.8
\times 10^{-4} (0.4~{\rm eV}/m_0)^2$, and using $\Delta m_{12}^2
\simeq 5 \times 10^{-5}$ eV$^2$ from solar neutrino oscillations, we
find $Re \delta'' < 7.8 \times 10^{-5} (0.4~{\rm eV}/m_0)^2$.  These
limits may be saturated mainly by $\tilde e_L - \tilde \tau_L$ mixing,
allowing $\tilde e_L - \tilde \mu_L$ mixing to be much more
suppressed.  In other words, from the data on neutrino oscillations,
we are able to make the direct connection in this model that flavor
violation in the charged-lepton sector should be the greatest in the
$\mu - \tau$ sector and smallest in the $e - \mu$ sector.

Using the same approximation which leads to Eq.~(26), the $\tau \to \mu
\gamma$ amplitude is calculated to be
\begin{equation}
{\cal A} = {e (3g_2^2 + g_1^2) \over 1536 \pi^2} (\sin \theta \cos \theta)
{m_\tau \over \tilde m_2^2} \epsilon^\lambda q^\nu \bar \mu \sigma_{\lambda
\nu} (1 + \gamma_5) \tau.
\end{equation}
The resulting branching fraction is then
\begin{equation}
B(\tau \to \mu \gamma) = 4.8 \times 10^{-6} \sin^2 \theta \cos^2 \theta
\left( {100~{\rm GeV} \over \tilde m_2} \right)^4.
\end{equation}
Using the experimental upper limit of $1.1 \times 10^{-6}$ on this
number, we require thus $\tilde m_2 > 102$ GeV.  Constraints from
$\tau \to e \gamma$ and $\mu \to e \gamma$ are also similarly satisfied.
Details will be given in a forthcoming comprehensive study of the
correlation between the neutrino parameters and the pattern of soft
supersymmetry breaking of this model.

Consider now the quark sector.  Whereas the neutrino sector has only
$L-L$ scalar mixings, we now also have $L-R$ and $R-R$ scalar mixings.
In a previous study \cite{bdm}, $V_{CKM} = 1$ was obtained from proportional 
$up$ and $down$ quark mass matrices, and it was shown that a realistic 
$V_{CKM}$ could then be generated with $L-R$ scalar quark mixings through 
gluino exchange.  Here $V_{CKM} = 1$ is obtained from our $A_4$ symmetry 
for any set of arbitrary $up$ and $down$ quark masses, with the obvious 
implication that the above result also applies.  [In the charged-lepton 
sector, the effect is smaller
and does not significantly change the neutrino mixing angles except
possibly for $U_{e3}$.]  More details will be discussed in the
forthcoming comprehensive study.

In conclusion, we have presented a concrete model based on the
discrete symmetry $A_4$ where quark and charged-lepton masses can be
all different and yet neutrino masses are degenerate at some high
scale where $V_{CKM} = 1$ and the effective neutrino mass matrix in
the $\nu_e, \nu_\mu, \nu_\tau$ basis is of the form
\begin{equation}
{\cal M}_\nu = \left[ \begin{array} {c@{\quad}c@{\quad}c} m_0 & 0 & 0 \\
0 & 0 & m_0 \\ 0 & m_0 & 0 \end{array} \right].
\end{equation}
The parameter $m_0$ naturally lies in the range where it can be probed
in cosmology, neutrinoless double beta decay and tritium beta decay.
Radiative corrections lift the neutrino degeneracy leading to (A)
$\sin^2 2 \theta_{atm} = 1$, (B) $U_{e3}$ small and imaginary,
and (C) large (but not maximal) solar mixing angle.  These corrections 
can be ascribed to
the structure of the soft supersymmetry breaking terms in the scalar
sector, which also break the $A_4$ symmetry explicitly and correlate
the neutrino mass matrix with lepton flavor violating processes.
Last but not least, a realistic quark mixing matrix $V_{CKM}$ may be
obtained in a totally analogous way.

~

\begin{center} {\bf Acknowledgements}\\
\end{center}

This work was supported in part by the U.~S.~Department of Energy
under Grants Nos.~DE-FG03-94ER40837, DE-FG03-98ER41076, and
DE-FG02-01ER45684, by a grant from the Research Corporation, by the 
European Commission Grant HPRN-CT-2000-00148, by the ESF
\emph{Neutrino Astrophysics
  Network}, and by the Spanish MCyT Grant PB98-0693.  The authors wish
  to thank particularly the Institute for Nuclear Theory, University of
  Washington, for its hospitality during the mini-workshop on
  neutrinos (April 2002) where this work was initiated.

\newpage
\bibliographystyle{unsrt}

\newpage

\begin{figure}
\begin{center}
\begin{picture}(270,110)(0,0)
\ArrowLine(0,50)(50,50)
\ArrowLine(270,50)(220,50)
\Line(50,50)(170,50)
\ArrowLine(170,50)(220,50)
\DashCArc(110,50)(60,0,180){4}
\Text(110,40)[]{$\tilde w$}
\Text(25,40)[]{$\nu_\mu$}
\Text(195,60)[]{$\nu_\tau$}
\Text(245,60)[]{$\nu_\mu$}
\Text(55,105)[]{$\tilde \mu_L$}
\Text(165,105)[]{$\tilde \tau_L$}
\Text(110,110)[]{$\times$}
\DashArrowLine(170,10)(220,50){4}
\DashArrowLine(270,10)(220,50){4}
\Text(170,0)[]{$\phi_2^0$}
\Text(270,0)[]{$\phi_2^0$}
\end{picture}
\end{center}
\caption{Wavefunction contribution to $\delta$ in supersymmetry.}
\end{figure}
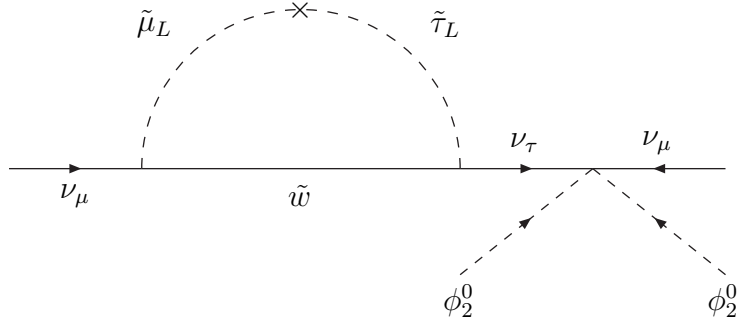

\begin{figure}
\begin{center}
\begin{picture}(270,110)(0,0)
\ArrowLine(0,50)(75,50)
\ArrowLine(270,50)(195,50)
\Line(0,50)(270,50)
\DashArrowLine(135,10)(135,50){4}
\DashArrowLine(195,10)(195,50){4}
\DashCArc(135,50)(60,0,180){4}
\Text(105,40)[]{$\tilde w$}
\Text(165,40)[]{$\tilde \phi_2$}
\Text(37,40)[]{$\nu_\mu$}
\Text(235,40)[]{$\nu_\mu$}
\Text(135,0)[]{$\phi_2^0$}
\Text(195,0)[]{$\phi_2^0$}
\Text(80,105)[]{$\tilde \mu_L$}
\Text(190,105)[]{$\tilde \tau_L$}
\Text(135,110)[]{$\times$}
\end{picture}
\end{center}
\caption{Vertex contribution to $\delta$ in supersymmetry.}
\end{figure}
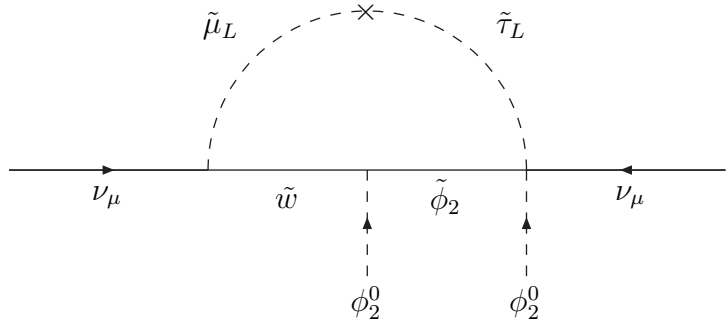

\end{document}